\documentclass[11pt,epsf]{article}
\usepackage{hyperref}
\usepackage{graphics}
\usepackage{amssymb}
\usepackage{epsf}
\usepackage{rotating}
\usepackage{pstricks}

\newcommand{\be}{\begin{equation}}
\newcommand{\ee}{\end{equation}}

\def\l{\lambda}

\def\ds#1{#1\kern-1ex\hbox{/}}
\def\dsh{h\kern-1.2ex /}

\newcommand{\bea}{\begin{eqnarray}}
\newcommand{\eea}{\end{eqnarray}}

\def\beq{\begin{equation}}
\def\eeq{\end{equation}}

\def\beqn{\begin{eqnarray}}
\def\eeqn{\end{eqnarray}}
\def\ba{\begin{eqnarray}}
\def\ea{\end{eqnarray}}

\newcommand{\beqa}{\begin{eqnarray}}
\newcommand{\eeqa}{\end{eqnarray}}

\begin{document}
\begin{center}
\vspace{1.cm}

{\bf \large Axion and Neutralinos from Supersymmetric Extensions of the Standard Model with anomalous $U(1)$'s\\}
\vspace{1.5cm}
{\bf $^{a}$Claudio Corian\`{o} $^{a}$Marco Guzzi, $^b$Nikos Irges and $^a$Antonio Mariano}

\vspace{1cm}
{\it $^a$Dipartimento di Fisica, Universit\`{a} del Salento \\
and  INFN Sezione di Lecce,  Via Arnesano 73100 Lecce, Italy\footnote{claudio.coriano@le.infn.it, marco.guzzi@le.infn.it, irges@physik.uni-wuppertal.de,antonio.mariano@le.infn.it}
}\\
\vspace{0.5cm}
{\it $^b$ Fachbereich C, Bergische Universit\"at Wuppertal \\
42097 Wuppertal, Germany\\}
\vspace{.5cm}
\vspace{.5cm}
\end{center}

\begin{abstract}
We analyze the conditions under which some supersymmetric 
generalizations of a class of models descending from string theory allow an axion in the physical spectrum, due to the 
presence of anomalous abelian gauge interactions. 
The gauge structure of these constructions involve the 
St\"uckelberg supermultiplet and a supersymmetric version of the
Wess-Zumino term for anomaly cancellation. While these conditions are 
not satisfied by the MSSM superpotential, we show that an axion-like 
particle appears in the spectrum if extra Standard Model singlets are present.
We show that the minimal requirements are met by simple superpotentials in which the singlet superfield is charged under the anomalous $U(1)$. 
The dark matter sector of these models include an axion and 
several neutralinos with an axino component.

\end{abstract}
\newpage

\section{Introduction}
One of the principal features of low energy effective string (inspired) models is the presence of a physical 
axion-like particle in their spectra. 
Models of this type (see for example \cite{Kiritsis:2003mc}, \cite{Svrcek:2006yi}) have received attention, given their simple gauge structure, which is typically characterized by extra abelian factors augmenting the 
gauge symmetry of the Standard Model (SM), some of which may be anomalous. These anomalous neutral currents are accompanied by axions for anomaly cancellation. The non-supersymmetric version of the low energy effective action for these theories has been investigated in detail in \cite{Coriano:2005js}, under the assumption that there is no decoupling of the anomalous interactions and of the corresponding anomalous gauge bosons, a situation which can be realized for example 
in scenarios with large extra dimensions.  
In this case the scale characterizing the breaking of the abelian 
symmetries - the St\"uckelberg mass $M_{S}$ - can be in the TeV region, opening the way for possible experimental signatures of these models at future colliders. In this work we discuss a variant of previous supersymmetric constructions of such models \cite{Anastasopoulos:2008jt}, based on the general discussion in the context of supergravity worked out in 
\cite{DeRydt:2007vg}, in order to generate a physical axion in their spectrum, which has not been found before. We summarize the salient steps, leaving the details to a forthcoming work.
We focus on the case of a single anomalous $U(1)$, here denoted as $U(1)_B$, where $B_\mu$ is the anomalous gauge boson, to differentiate it from the hypercharge of the SM, denoted as  $U(1)_Y$. In effective string models this simple abelian structure can be made more general with the introduction of several $U(1)$ factors, described in the hypercharge basis by direct products of the form 
$G_1\equiv U(1)_Y\times U(1)_1\times ... \times U(1)_p$, with an anomaly-free hypercharge generator and $p$ 
anomalous $U(1)$'s which are accompanied by axions $b_i$, with $i=1,2,...p$. The anomalous $U(1)$'s in this construction are in a broken phase, dubbed the ``St\"uckelberg phase". After electroweak symmetry breaking 
the massive neutral gauge bosons acquire their mass via a combination of the Higgs and of the St\"uckelberg mechanisms. 
Due to the anomalous nature of these models, the role of the St\"uckelberg fields is enhanced compared to 
just a simple modification of the symmetry breaking mechanism, because of the presence of axion couplings of these fields to the anomaly, which may induce a physical axion in the spectrum. 
The identification of this state is rather subtle due to the combination of the two mechanisms, and therefore complete information on this comes from both the Higgs potential and by an analysis of the bilinear mixings $Z_i\partial G_{Z_i}$. The latter are crucial for the identification of the goldstone modes of the neutral gauge fields $G_{Z_i}$ in the Higgs-St\"uckelberg phase.
\label{Intro}

\section{Higgs-axion mixing}

In order to briefly highlight the property of Higgs-axion mixing in these models, we recall that the typical Higgs potential that appears in the non-supersymmetric analysis of these models  \cite{Coriano:2005js} is the most general $SU(2)_L\times G_1$ invariant constructed from the two Higgs $SU(2)$ doublets $H_u$ and $H_d$ of charge $q_u$ and $q_d$ under $U(1)_B$ \footnote{The example below is in a basis where $H_u$ and $H_d$ have equal hypercharge. This is not the usual convention of the MSSM, however is the one that is more convenient for our discussion. It is not hard to transform into the usual basis, see the Appendix of  \cite{Coriano:2005js}.}
\beq
V=\sum_{a=u,d}\Bigl(  \mu_a^2  H_a^{\dagger} H_a 
+ \l_{aa} (H_a^{\dagger} H_a)^2\Bigr)
-2\l_{ud}(H_u^{\dagger} H_u)(H_d^{\dagger} H_d)
+2{\l^\prime_{ud}} |H_u^T\tau_2H_d|^2 \label{PQ},
\eeq
while the presence of shifting axions allows for a non-holomorphic sector
\begin{eqnarray}
V^\prime&=& \lambda_0\, H_u^{\dagger}H_d e^{-i(q_u-q_d)\frac{b}{M_S}}
+\l_1 \left(H_u^{\dagger}H_d e^{-i(q_u-q_d)\frac{b}{M_S}}\right)^2  
\nonumber\\
&&+\l_2 \left(H_u^{\dagger}H_u\right)
H_u^{\dagger}H_d e^{-i(q_u-q_d)\frac{b}{M_S}}
+\l_3 \left(H_d^{\dagger}H_d\right)
H_u^{\dagger}H_d e^{-i(q_u-q_d)\frac{b}{M_S}} + c.c.
\label{PQbreak}
\end{eqnarray}
This second sector guarantees the presence of a physical axion in the spectrum. 
In particular, the first coupling in this second sector is the one that corresponds to the term in
the MSSM potential (where it would appear without the phase of course) which gives mass to $A^0$. 
In this basis it triggers the mixing of the
axion to the Higgs and also contributes to its mass, as do the rest of the terms. 

To see this, we parameterize the Higgs fields in terms of 8 real degrees of freedom as
\begin{eqnarray}
H_u=\left(\begin{array}{c}
H_u^+\\
H_u^0
\end{array}\right) \qquad H_d=\left(\begin{array}{c}
H_d^+\\
H_d^0 
\end{array}\right)
\end{eqnarray}
where $H_u^+$, $H_d^+$ and $H_u^0$, $H_d^0$ are complex fields. We focus on the neutral CP-odd sector. In this case, expanding  around the vacuum we get for the uncharged components
\beq
H_u^0 =  v_u + \frac{\textrm{Re}H_{u}^0 + i \textrm{Im}H_{u}^0}{\sqrt{2}} , \qquad
H_d^0 =  v_d + \frac{\textrm{Re}H_{d}^0 + i \textrm{Im}H_{d}^0}{\sqrt{2}}, 
\label{Higgsneut}
\eeq
with quadratic contributions given by
\beqa
V_{CP-odd}&=&
 \left( \textrm{Im} {H_u}^0,  \textrm{Im}{H_d}^0, b \right){\cal N}_3\left(\begin{array}{c}
 \textrm{Im}{H_u}^0\\
 \textrm{Im}{H_d}^0 \\
b\\
\end{array}\right)
\eeqa
for a suitable ${\cal N}_3$ matrix.
The bilinear mixing terms $Z \partial G_Z$ which 
allow to identify the goldstones of the massive (physical) gauge bosons are extracted from the kinetic terms
\beq
|{\cal D}_\mu H_u|^2 + |{\cal D }_\mu H_d|^2 +
\frac{1}{2}(\partial_\mu b+ M_{S} B_\mu)^2,\label{quadform}
\eeq
where $b$ is the axion field.
By means of an orthogonal rotation one can relate the mass to the interaction eigenstates according to
\be
\pmatrix{{\rm Im}H_u^0\cr {\rm Im}H_d^0\cr b}=\; O\;
\pmatrix{\chi \cr G_1^0\cr G_2^0}\label{rotunit}
\ee
with $O$ being an orthogonal matrix. We have denoted the physical field by $\chi$ and the  NG-bosons
by $G_{1,2}^0$. 
If the eigenvectors corresponding to the interaction basis states are substituted back in the scalar potential
one should obtain mass terms for the physical fields (except for those that happen to be massless) while the quadratic terms corresponding to unphysical states should vanish identically upon imposing the ``vacum condition", just like in the Standard Model. Clearly this imposes a severe constraint on possible models that hope to generate a physical axion, especially for theories with scalar potentials restricted by larger symmetry (like supersymmetry).

In the presence of $V^\prime$, $\chi$ acquires a physical axion-like coupling and becomes a massive 
axion. For a potential such as the one given in eq. (\ref{PQ}) instead, the St\"uckelberg axions introduced to render the extra $U(1)$'s massive are merely goldstone modes. In this case from the 
unphysical bilinears one identifies only one physical CP-odd Higgs (called $A^0$ in the MSSM) which can not have an axion-like coupling as can be verified by also a simple counting of the degrees of freedom before and after electroweak symmetry breaking. 
In this case eq. (\ref{rotunit}) simplifies and takes the form
\bea
\pmatrix{{\rm Im}H_u^0\cr {\rm Im}H_d^0}=\; O\;
\pmatrix{A^0 \cr G^0}
\label{rotunit2}
\eea
and it is clear that the physical state in the CP-odd sector does not acquire an axion coupling.
The above discussion carries through in a similar way for the anomalous $U(1)$ extension of the MSSM \cite{Anastasopoulos:2008jt}.

One should however realize that this situation is not generic. 
In fact, there are cases in which even in the absence of 
a direct coupling of $b$ to $H_{u,d}$ one can still have a physical axion in the spectrum. 
We are going to describe below 
how to obtain in a class of supersymmetric models a massless CP-odd scalar that acquires an axion-like coupling. 
After supersymmetry breaking, terms of the type $V^{\prime}\;$ may be induced, making the axion massive.
The induced mass by these terms may be tiny and up to the electroweak breaking scale, depending on the 
couplings parametrizing $V^{\prime}\;$ \cite{Coriano:2005js}, \cite{Coriano:2006xh}, \cite{Coriano:2007xg}.
\label{sec1}

\section{Supersymmetric models with axion-like particles} 

We first recall that supersymmetric extensions \cite{Anastasopoulos:2008jt} of the class of models introduced, besides the usual supersymmetric gauge multiplets for the 
 $SU(3)\times SU(2)\times U(1)_Y\times U(1)_B$ gauge symmetry, the St\"uckelberg multiplet \cite{Kors:2004ri}
\begin{eqnarray}
{\cal L}_{S}=\int d^{4}\theta\left[2 M_{S}\hat{B}+\hat{{\bf b}}+\hat{{\bf b}}^{\dagger} \right]^{2}
\end{eqnarray}
where $\hat{B}$ is the abelian scalar superfield associated to the extra $U(1)_{B}$ and
$\hat{{\bf b}}$ is a left-chiral superfield. St\"uckelberg fields have been the subject of interesting phenomenological studies \cite{Feldman1}. The physical components 
of $\hat{\bf b}$ are the complex axion $b$ and its supersymmetric partner, the axino $\psi_{\bf b}$.
To this action one adds the WZ counterterms which represent the supersymmetric Wess-Zumino interaction invoked 
for anomaly cancellation, with a counterterm lagrangian given by \cite{Anastasopoulos:2008jt}\newpage
\begin{eqnarray}
\hspace{-1cm}
{\cal L}_{WZ}&=&-\int d^{4}\theta\left\lbrace\left[
\frac{1}{2} b_{G} \,\textrm{Tr}({\cal G} {\cal G})\hat{{\bf b}}
+
\right.\right.
\left.\left.
\frac{1}{2} b_{W} \,\textrm{Tr}(W W)\hat{{\bf b}} +b_{Y}\hat{{\bf b}}W^{Y}_{\alpha} W^{Y,\alpha}
+b_{B}\hat{{\bf b}}W^{B}_{\alpha}W^{B,\alpha}
\right.\right.
\nonumber\\
&&\left.\left.
\hspace{3.5cm} +b_{YB}\hat{{\bf b}}W^{Y}_{\alpha}W^{B,\alpha}\right]
\delta(\bar{\theta}^{2})
+h.c.\right\rbrace.
\end{eqnarray}
In order to identify a supersymmetric model which has such characteristics, we choose a
superpotential in which the $\mu$ term  ($\mu \hat{H}_{u} \cdot \hat{H}_d$ ), which is present in the MSSM, is generated by the vev of an extra singlet superfield 
(from now on we switch to the conventional vector-like hypercharge assignment of the MSSM Higgses)
\begin{eqnarray}
\lambda\hat{S}\hat{H}_{u}\cdot\hat{H}_{d}
\label{super}
\end{eqnarray}
and with a gauge structure which is enlarged with a single anomalous $U(1)$.  The extra singlet superfield has one bosonic and one fermionic component $(S, \tilde{S})$ respectively, with a complex $S$. We choose a charge assignment in such a way that $\hat{S}$ is a SM singlet, but carries charge under the anomalous $U(1)$. 
Notice that we are not allowing a linear $\hat{S}$ or a cubic ($\hat{S}^3$) in the superpotential. The first case is contemplated by the nMSSM, while the second case is that of the NMSSM. Therefore this superpotential is similar to that of the USSM previously considered in
\cite{Langacker:2008yv,Cvetic:1997ky,Suematsu:1994qm}.
It can be shown that this charge assignment 
can be arranged in order to cancel all the additional anomalies induced by the extra abelian symmetry. 
From our discussion in the previous section this term can be recognized as an analogue of the first term
in eq. (\ref{PQbreak}) which in the light of eq. (\ref{super}) can be viewed as a term of this type, with its radial fluctuation frozen perhaps from some spontaneous symmetry breaking at a higher scale. Even though the axion is not 
the phase of $S$,
the addition of this extra SM singlet superfield is sufficient to remove the second massless mode in the mass matrix of the neutral gauge bosons, while it provides the necessary enlargement of the CP-odd sector so that room for 
a physical axion appears. 
 
The axion is then searched in the linear combination
\ba
\chi=b_{1}\textrm{Im}~H_u^0+b_{2}\textrm{Im}~H_d^0+b_{3}\textrm{Im}~S+b_{4}\textrm{Im}~b,
\ea
and is found to be \cite{new_pap:00}
\ba
&&\hspace{-1cm}
\chi=
\frac{1}{N_{\chi}}\left[2 M_{S}v_u v_d^2\, \textrm{Im}H^0_u + 2 M_{S}v_u^2 v_d\, \textrm{Im}H^0_d 
\right.
\left.
-2 M_{S}v^2 v_S \,\textrm{Im}\,S + q_S\,g_B (v^2 v_S^2 + v_u^2 v_d^2)\textrm{Im}\,b\right]\,,
\nonumber\\
&&\hspace{-1cm}
N_{\chi}=\sqrt{4 M_{S}^2 v^2 (v^2 v_S^2 + v_u^2 v_d^2)+q_S^2 g_B^2 (v^2 v_S^2 + v_u^2 v_d^2)^2}
\ea
where $v=\sqrt{v_u^2+v_d^2}$ and $v_S$ are the vevs that $H_{u,d}$ and $S$ take from the scalar potential
\begin{eqnarray}
V&=&\vert\lambda H_{u}\cdot H_{d}\vert ^{2}
+\vert\lambda S\vert ^{2}(\vert H_{u}\vert^{2}+\vert H_{d}\vert^{2})
+\frac{1}{8}(g_2^{2}+g_Y^{2})(H_{u}^{\dagger}H_{u}-H_{d}^{\dagger}H_{d})^{2}\nonumber\\
&&+\frac{g_B^{2}}{8}(q_{H_{u}}H_{u}^{\dagger}H_{u}
+q_{H_{d}}H_{d}^{\dagger}H_{d}+q_{S}S^{\dagger}S)^{2}
+\frac{g_2^{2}}{2}\vert H_{u}^{\dagger}H_{d}\vert^{2}
+m_{1}^{2}\vert H_{u}\vert^{2}+m_{2}^{2}\vert H_{d}\vert^{2}\nonumber\\
&&+m_{S}^{2}\vert S\vert^{2}+(a_{\lambda}SH_{u}\cdot H_{d}+h.c.).
\end{eqnarray}
The CP-odd sector, in this case, can be spanned by the goldstones $G_Z,G_{Z^{\prime}}$ and the physical directions $A^0$ and $\chi$. 
The rotation matrix $O$ is defined as
\ba
\hspace{1cm}
\left(
\begin{array}{c}
A^{0}\\
G_Z\\
G_{Z'}\\
\chi
\end{array}
\right)=
O
\left(
\begin{array}{c}
\textrm{Im}~H_u^0\\
\textrm{Im}~H_d^0\\
\textrm{Im}~S\\
\textrm{Im}~b
\end{array}
\right).
\ea
We have therefore shown that even in the absence of direct mixing of the St\"uckelberg axion with the Higgs in the potential, a massless physical axion is still allowed by the theory.

\subsection{Other cases: the $U(1)$ extension of the nMSSM} 

The attempt to generalize this analysis to other supersymmetric cases clarifies once more that the presence or the absence of a massless physical 
CP-odd scalar with an axion coupling, in the absence of direct mixing, is related to the structure of the superpotential. 

As a further example, let us consider the case of the superpotential of the nMSSM, with a gauge symmetry which is extended with an extra $U(1)$ \cite{Langacker:2008yv}. In our case we assume this $U(1)$ to be anomalous.
This superpotential contains an extra linear term in 
$\hat{S}$ \cite{tamvakis}  respect to eq. (\ref{super}), which is allowed by the gauge symmetry 
since the $U(1)$ charge of the singlet is vanishing $(q_S=0)$. In this case the charge assignment of the two Higgses clearly has to be $q_{H_{u}}=-q_{H_{d}}$. 
As usual, the rigorous identification of the physical states needs a joint analysis of the CP-odd sector of the potential and of the bilinear mixings, even though the absence of  a physical axion due to the decoupling of $\hat S$ from the Higgses 
can be already anticipated.
Specifically, in the basis $\{\textrm{Im}\,H^{0}_u,\textrm{Im}\,H^{0}_d, \textrm{Im}\,S\}$ we find a single goldstone 
and two physical states. The goldstone mode identified from the Higgs potential is given by

\beq
G^{0}=\frac{1}{v}\left(v_u \textrm{Im}\,H^{0}_u-v_d \textrm{Im}\,H^{0}_d\right).
\eeq
This analysis would be sufficient to reach the conclusion that there is no pseudoscalar with an axion-like coupling in the spectrum. In fact, the complete CP-odd sector is spanned by 4 states, two of which have been identified from the 
Higgs potential. We remark that the goldstone of the Higgs potential belongs to the subspace spanned by the two (true) goldstones, which are identified by a parallel analysis of the bilinear mixings. 
These are given as a linear combination of $\textrm{Im}\,b$ and 
of $G^{0}$ 
\ba
G_Z= \alpha_1 G^{0} + \alpha_2 \textrm{Im}\,b ,
&&
G_{Z'}= \alpha'_1 G^{0} + \alpha'_2 \textrm{Im}\,b.
\ea
Equivalently (and more simply), using the quadratic bilinear mixings we can span the whole 
CP-odd sector using the basis 
$(G_Z,G_{Z^{\prime}},H_1, H_2)$, where we have denoted with $H_{1,2}$ the two physical Higgs eigenstates 
(which should coincide of course with the ones obtained from the potential). At this point, since the dimension of the sector is 4 and the physical scalars have to be orthogonal 
to the subspace spanned by the two true goldstones, the two orthogonal directions are indeed physical, but they have to accomodate the two physical Higgses of the sector. These are deprived of axion-like couplings - being 
pure Higgs states - and the model, therefore, does not allow a physical axion in its pseudoscalar spectrum.

\subsection{Multiple $U(1)$ and multiple Higgs models}

The generalization of our considerations to the case of more complex models 
(such as of those coming from intersecting branes for example) 
is quite straightforward. In this case, the extra abelian symmetries may appear in the effective description already in a broken phase, and may be anomalous or anomaly-free.
We assume for simplicity that each axion shifts under only one $U(1)$.
As usual, St\"uckelberg mass terms can combine with the electroweak symmetry breaking mechanism to give masses to the extra $U(1)$'s. Also in this case we choose a renormalizable 
superpotential which is of the type given in eq. (\ref{super}) with an extra SM singlet, here denoted as $\hat{S}_j$. The 
index $j$ selects the corresponding anomalous $U(1)_j$  under which the charge of the singlet is nonzero. In other words we assume for simplicity that $\hat{S}_j$ is an overall singlet, except for its charge under the the j-th anomalous $U(1)$. It is then obvious that the previous analysis can be extrapolated to this case as well. We perform a combined analysis of the potential and of the bilinear mixings. We may allow combined 
St\"uckelberg and Higgs charges for all of the $U(1)'s$. 

The lagrangian which gives the contribution to the mass of the gauge bosons
is given by
\ba
{\cal L}_{\textsc{quad}}&=&\vert{\mathcal D}_{\mu}H_{u}\vert^{2}
+\vert{\mathcal D}_{\mu}H_{d}\vert^{2}
+\vert{\mathcal D}_{\mu}S_j\vert^{2}
+\frac{1}{2}\left(\partial_{\mu}\textrm{Im}~b_j + M_{S_j} B^{(j)}_{\mu}\right)^2
 + \sum_{i\neq j} \frac{1}{2}\left(\partial_{\mu}\textrm{Im}~b_i + M_{S_i} B^{(i)}_{\mu}\right)^2 \nonumber \\
\ea
and involves, besides the two Higgses, the SM bosonic singlet of $\hat{S}_{j}$, the bosonic component of 
the St\"uckelberg axions $b_{i,j}$ and the St\"uckelberg masses $M_{S_{i,j}}$. It is then clear that the axions $b_i$ ($i\neq j$) are goldstone modes, while the potential will generate a mixing between $\textrm{Im} H_u,\textrm{Im} H_d$ and 
$\textrm{Im} S_j$, with a single CP-odd physical state $A^0$ and two goldstone modes. 
The  identification of the physical axion 
can be easily performed in the subspace spanned by $(G_Z, G_{Z_j}, \textrm{Im} \,b_{i\ne j}, \textrm{Im} \,b_{j})$, with the physical axion 
$\chi$ identified by the eigenstate which is orthogonal to the subspace spanned by $G_Z, G_{Z_j}$ and $A^0$. It is quite obvious that if we take $\hat{S_j}$ to be a singlet of the entire gauge symmetry, then axion couplings are not allowed. 
 
The generalization of this analysis to more general Higgs structures now becomes a simple application of Goldstone's Theorem.  We consider a model with $p$ anomalous abelian symmetries
each with its own St\"uckelberg axion
and $n_H$ CP-odd components in the Higgs sector. We also assume to have $n_S$ singlets which couple to the $p$ abelian symmetries. The dimension of the CP-odd sector is $n_{CP}\equiv n_H + n_S + p$. We 
require that only the $n_S$ singlets couple to the Higgs sector, together with the property that from the Higgs potential 
we isolate $n_{\textrm{phys}}$ physical components. Then the condition to be satisfied in order to have a physical axion is that the subspace spanned by the $p+1$ goldstones of the broken gauge bosons, corresponding to $Z, Z_1,...,Z_p$, together with the $n_{\textrm{phys}}$ physical states identified by the Higgs potential, leave additional space for one physical direction in the CP-odd sector 
$p+1 + n_{phys} = n_{CP} -1$. This additional direction completely defines an axion-like component in this 
sector. 
\begin{figure}[h]
\begin{center}
\includegraphics[width=6cm,angle=-90]{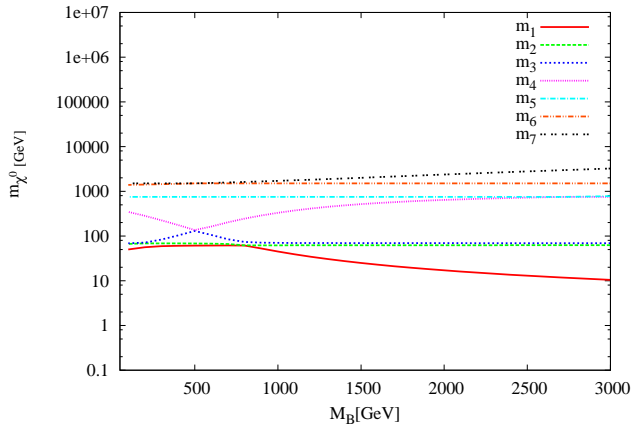}
\caption{\small Spectrum of the neutralino sector in the model}
\label{neutral}
\end{center}
\end{figure}

A last comment concerns the neutralino sector in these models in the $U(1)$ extension described by the choice of the superpotential in eq. (\ref{super}). In this case the neutralino sector contains seven states, built out of the higgsinos 
$( \lambda_{W^3},\lambda_{Y})$, the Bino ($\lambda_{B}$) which is the susy partner of the anomalous gauge boson $B_\mu$, the two higgsinos $(\tilde{H_{u}^{1}},\tilde{H_{d}^{2}})$, the singlino ($\tilde{S}$), besides the axino $\psi_{\bf b}$. We show in Fig. \ref{neutral} plots of the ordered neutralino eigenvalues of the model as a function of the gaugino mass 
term $M_B$ of the anomalous gauge boson for a model with a single anomalous $U(1)$.  
We have chosen a coupling constant 
$g_B=0.65$ and selected a representative value of the St\"uckelberg mass $M_{S}=3$ TeV with $\tan\beta =40$. 
The other values of the soft breaking parameters of the model that we have chosen 
are $M_Y=1.5$ TeV for the $U(1)_{Y}$ gaugino mass term, $M_w=3$ TeV
for $SU(2)$, $M_{\bf b}=3$ TeV for the axino mass term.
The doublet-singlet mixing parameter is $\lambda=0.1$ while for the Higgs and for the
extra singlet vevs we have chosen $v_u\approx 6$ GeV and $v_S\approx 1$ TeV.
The charge assignment for $\{q_{H_u},q_{H_d},q_S\}$ is 
$\{-3/(2\sqrt{10}),-2/(2\sqrt{10}),5/(2\sqrt{10})\}$.
For details, see \cite{new_pap:00}.

\section{Conclusions}
Recent studies of a class of vacua of string theory have addressed at a certain level of detail the issue of anomalous abelian symmetries. Their supersymmetric extensions, developed using a bottom-up approach, have the goal of identifying the key phenomenological features and implications of anomalous $U(1)$'s in effective models. We have shown 
that simple superpotentials can produce a pseudoscalar in the spectrum. The absence of Higgs-axion mixing terms due to the requirement of holomorphicity does not forbid a massless CP-odd state in these models with a
Wess-Zumino coupling.
Furthermore, supersymmetry breaking induces in general Higgs-axion mixing terms in the scalar potential, which can give a mass to the axion
that can reach up to the electroweak breaking scale, apart from the usual instanton contribution.
In view of the more recent  attention towards the detection of pseudoscalars in forthcoming experiments \cite{Ahlers:2007qf,Ahlers:2007rd,roncadelli1,roncadelli2}, the study of these models is likely to receive a further boost in the near future. 
At the same time one can envision the possibility for realistic applications of our results to cosmology as in the MSSM case \cite{Enqvist}, such as in the study of modular inflation \cite{Dimopoulos:2005bp} and similar aspects in cosmology where the role of pseudoscalars is of possible importance. Here we have shown that this particle can be gauged and obtain a physical status in models which are fully compatible with supersymmetry.

\centerline{\bf Aknowledgements}
The work of C.C. was supported in part
by the European Union through the Marie Curie Research and Training Network ``Universenet'' (MRTN-CT-2006-035863) and by the Interreg II Crete-Cyprus program. 
The work of N.I. is supported by a Research Fellowship for Experienced Researchers of the
Alexander von Humboldt Foundation.

\end{document}